\documentclass[journal=aamick,manuscript=article]{achemso}
\usepackage{chemformula} 
\usepackage[T1]{fontenc} 
\usepackage[locale=UK, exponent-product = \cdot, separate-uncertainty=true,list-units=single,range-units=single]{siunitx}
\usepackage{amssymb}

\DeclareSIUnit{\mymmol}{m\textsc{m}}

\newcommand*{\ind}[1]{_\text{#1}}
\newcommand*{\Du}{\text{Du}}

\author{Alexander Kiy}
\email{alexander.kiy@anu.edu.au}
\author{Shankar Dutt}
\author{Christian Notthoff}
\affiliation[ANU]{Department of Materials Physics, Research School of Physics, Australian National University, Canberra ACT 2601, Australia}
\author{Maria E. Toimil-Molares}
\affiliation[GSI]{GSI Helmholtzzentrum für Schwerionenforschung, Planckstr. 1, 64291 Darmstadt, Germany}
\author{Nigel Kirby}
\affiliation[Synchrotron]{Australian Synchrotron, ANSTO, 800 Blackburn Rd, Clayton VIC 3168, Australia}
\author{Patrick Kluth}
\affiliation[ANU]{Department of Materials Physics, Research School of Physics, Australian National University, Canberra ACT 2601, Australia}

\title {Highly Rectifying Conical Nanopores in Amorphous \ch{SiO2} Membranes for Nanofluidic Osmotic Power Generation and Electroosmotic Pumps}

\begin{document}

\begin{abstract}
    Nanopore membranes are a versatile platform for a wide range of applications ranging from medical sensing to filtration and clean energy generation. To attain high-flux rectifying ionic flow, it is required to produce short channels exhibiting asymmetric surface charge distributions. This work reports on a system of track etched conical nanopores in amorphous \ch{SiO2} membranes, fabricated using the scalable track etch technique. Pores are fabricated by irradiation of \SI{920(5)}{\nm} thick \ch{SiO2} windows with \SI{2.2}{\GeV} \ch{^{197}Au} ions and subsequent chemical etching. Structural characterisation is performed using atomic force microscopy, scanning electron microscopy, small-angle X-ray scattering, ellipsometry, and surface profiling. Conductometric characterisation of the pore surface is performed using a membrane containing 16 pores, including an in-depth analysis of ionic transport characteristics. The pores have a tip radius of \SI{5.7(1)}{\nm}, a half-cone angle of \SI{12.6(1)}{\degree}, and a length of \SI{710(5)}{\nm}. The $\text{p}K_a$, $\text{p}K_b$, and pI are determined to \num{7.6(1)}, \num{1.5(2)}, and \num{4.5(1)}, respectively, enabling the fine-tuning of the surface charge density between +100 and \SI{-300}{\milli\coulomb\m\tothe{-2}} and allowing to achieve an ionic current rectification ratio of up to 10. This highly versatile technology addresses some of the challenges that contemporary nanopore systems face and offers a platform to improve the performance of existing applications, including nanofluidic osmotic power generation and electroosmotic pumps.
\end{abstract}
\section{Introduction}
Over the past two decades, nanopore technology has developed into a versatile platform for a wide range of applications such as ultrafiltration, biosensing and medical sensing, nanofluidics, and nanoelectronic devices.\cite{Dekker2007,Ahn2014,Lemay2009,Venkatesan2011,Luan2019} Solid-state nanopores are robust and durable under harsh conditions such as elevated temperature, extreme pH environments, and high pressure. They are highly tunable in geometry and surface properties and thus enable novel applications. Examples include filtration of water to remove contaminants such as oil\cite{Gao2014} and dyes,\cite{Li2020} nanopore sensors for single molecules,\cite{Liu2021} proteins,\cite{Houghtaling2018} blood sugar,\cite{Galenkamp2018} and drugs\cite{Wang2018} as well as DNA,\cite{Wang2021a} ssDNA,\cite{Zhu2022} and protein sequencing.\cite{Yu2019a} The potential of using nanofluidic devices based on nanopore membranes for novel applications such as power generation\cite{Laucirica2021} and electroosmotic pumps\cite{Wu2016} has also been explored in detail.

Solid-state nanopore membranes can be manufactured in a wide range of materials such as semiconductors,\cite{Li2001,Storm2003,Zhang2007,Kluth2008b} polymers,\cite{ToimilMolares2001,Fink2,Kiy2021,Wang2022} alumina,\cite{Yuan2004} carbon,\cite{Zhan2020} graphene,\cite{Chen2022} borophene,\cite{Jena2021} and \ch{MoS2}\cite{Ke2019} with only a single\cite{Chtanko2004} or multiple\cite{Harrell2006,Scopece2006} pores depending on the material and fabrication technique. One major challenge in current nanopore technology is to combine the high throughput rates that nanopores in ultrathin membranes provide with highly asymmetric transport properties characteristic of long, conical nanopores.\cite{PerezSirkin2020,Tang2016} Furthermore, translating the performance of single pores to multipore systems without losing performance has proven to be challenging.\cite{Cao2018,Laucirica2021}

Different techniques like e-beam lithography,\cite{Deshmukh1999} ion beam sculpting,\cite{Li2001} e-beam drilling using a transmission electron microscope,\cite{VanDenHout2011} focused ion beam drilling,\cite{Tang2014} dielectric breakdown,\cite{Goto2019} or laser-assisted pulling\cite{Cadinu2020} can only be used to fabricate a single or few nanopores and thus lack the scalability that many applications require. Additionally, these techniques do not allow to precisely shape the geometry of the pores. Ion track etching has been shown to create versatile conical nanopores in polymers\cite{Li2004} and silicon nitride\cite{Vlassiouk2009} in a way that is industrially scalable. In this work, we present a system of conical nanopores in thin amorphous \ch{SiO2} (a-\ch{SiO2}) membranes fabricated using ion track etching. Our membranes feature conical nanopores that exhibit opening angles far greater than those in track etched polymers,\cite{Kaya2016,Hadley2020} while being an order of magnitude thinner. We can manufacture either single pores or multipores with adjustable pore density on large areas. As a-\ch{SiO2} is widely used in the semiconductor industry, the fabrication techniques are precise and cost-efficient and can be readily integrated in lab-on-a-chip devices.

Here, we present the fabrication of multipore membranes, and the detailed characterisation of the structure and ionic transport properties. The potential performance gain of implementing this nanopore platform in an electroosmotic pump or nanofluidic osmotic power generation system in comparison to existing nanopore membranes is explored.

\section{Experimental Section}
\subsection{a-\ch{SiO2} Membrane Fabrication}
\begin{figure}
	\centering
	\includegraphics[width=.49\textwidth]{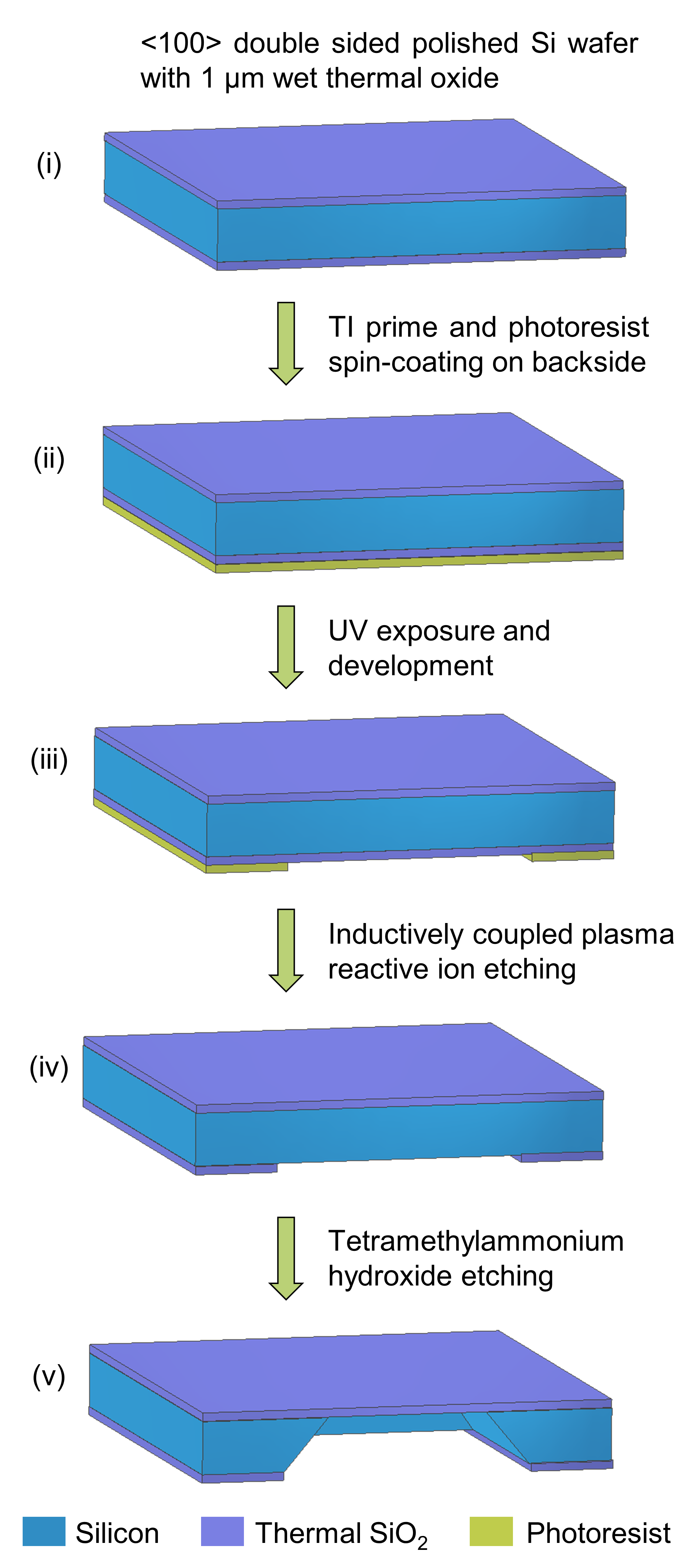}
	\caption{Fabrication process of a-\ch{SiO2} membranes. (i) Si wafer with a \SI{920}{\nm} layer of \ch{SiO2} on each side is (ii) coated with photoresist on the backside. (iii) Using UV lithography, a pattern for the membranes is created, (iv) which is followed by reactive ion etching. (v) Finally, tetramethylammonium hydroxide etching removes the silicon selectively to \ch{SiO2}, and after RCA cleaning, fabrication of the a-\ch{SiO2} membranes is completed.}
	\label{f:fab}
\end{figure}

Figure~\ref{f:fab} shows the workflow involved in the fabrication of silicon dioxide membranes. Double side polished $\langle$100$\rangle$ silicon wafers of \SI{100}{\mm} diameter with \SI{920(5)}{\nm} wet thermal silicon dioxide were purchased from WaferPro, LLC (USA) (i). The wafers were then RCA cleaned before being spin-coated with a thin coating of TI prime (Microchemicals, GmbH) for \SI{25}{\second} at \SI{3000}{\text{rpm}}, which acts as an adhesion promoter. On the backside of the wafer, negative photoresist maN-1420 (Microchemicals, GmbH) was spin-coated for \SI{30}{\second} at \SI{3000}{\text{rpm}} (ii). After that, UV lithography was used to pattern custom windows measuring $475\times\SI{475}{\um\tothe{2}}$ (iii). The silicon was then exposed in the customised window region by removing the a-\ch{SiO2} layer using reactive ion etching from the backside of the wafer (iv). The photoresist was removed, and the exposed \ch{Si} was then anisotropically etched using a wet etching solution containing \SI{5}{\%} tetramethylammonium hydroxide (TMAH) (Sigma-Aldrich, 331635) which leads to the formation of membranes (v). TMAH has a higher selectivity of a-\ch{SiO2}/\ch{Si} than other etchants like potassium hydroxide, and the exposed \ch{Si} can be completely etched without the need for extra silicon nitride layers. The membranes were then RCA cleaned to eliminate any contaminants they may have acquired during the process.  On a \SI{100}{\mm} Si wafer, this method results in the creation of 220 a-\ch{SiO2} membranes with a frame size of $5.6\times\SI{5.6}{\mm\tothe{2}}$ and a window size of $55\times\SI{66}{\um\tothe{2}}$.

\subsection{Theoretical Background}
The total conductance of a membrane $G\ind{m}$ with $n$ conical nanopores is the sum of the conductance of each individual nanopore $G\ind{p}$, given by $G\ind{m}=\sum_n G\ind{p}=n G\ind{p}$. As the average distance between pores is on the order of \SI{10}{\um}, inter-pore effects can be neglected.\cite{Cao2018} The pore conductance consists of the bulk conductance $G\ind{b}$, the surface conductance $G\ind{s}$, and the access resistance at the cone tip $R\ind{a,t}$ and base $R\ind{a,b}$:\cite{Hall1975,Lee2012}
\begin{equation}
    G\ind{m}=n\left(\frac{1}{G\ind{b}+G\ind{s}}+R\ind{a,t}+R\ind{a,b}\right)^{-1}\label{eq:cond}
\end{equation}

As charged conical nanopores often exhibit asymmetric transport properties between positive and negative applied voltage, we analysed the conductance $G_0$ in the vicinity of \SI{0}{\V} where it is largely linear in a small bias interval. We determined the conductance in the linear regime of the $I-V$ curve around \SI{0}{\V} between \SIlist{-.1;.1}{\V}. The different contributions are\cite{Frament2012}

\begin{align}
    G_{\text{b},0}&=\kappa\ind{b}\frac{\pi r\ind{t} r\ind{b}}{L}\\
    G_{\text{s},0}&=\kappa\ind{b}\frac{r\ind{b}-r\ind{t}}{\ln\left(\frac{r\ind{b}}{r\ind{t}}\right)}\frac{\pi}{L}\frac{\left|\sigma\right|}{e c}
\end{align}
where $\kappa\ind{b}$ is the bulk conductivity, $e$ is the elementary charge, $c$ is the salt concentration, $r\ind{t}$ is the radius at the cone tip, $r\ind{b}$ is the radius at the cone base, $L$ is the pore length, $\sigma$ is the surface charge density, and $\Du_i=\left|\sigma\right|/(e c r_i)$ is the Dukhin number at the pore orifices.

The access resistance according to Hall is given by\cite{Hall1975,Lee2012}
\begin{equation}
    R_{\text{a},i}=\frac{1}{\kappa\ind{b} r_i}\frac{1}{4+\Du_i}
\end{equation}
and describes the rate at which the ions pass into and out of the nanopore at the cone tip and base.

In a-\ch{SiO2} nanopores, the surface charge is regulated by protonation and deprotonation reactions of silanol (\ch{SiOH}) groups at the pore surface that can render the surface negatively or positively charged:\cite{Yang2020}

\begin{align*}
    \ch{SiOH &<=>[$K\ind{a}$] SiO- + H+}\\
    \ch{SiOH + H+ &<=>[$K\ind{b}$] SiOH2+}
\end{align*}
with the acid and base dissociation constants $K\ind{a}$ and $K\ind{b}$ that are related to the $\text{p}K_i$ of the surface $K_i=10^{-\text{p}K_i}$. The surface charge density $\sigma$ is controlled by the local $\zeta$-potential:\cite{Smeets2006,Frament2012}
\begin{equation}
    \zeta(\sigma)=-\frac{k\ind{B}T}{e}\ln\left(\frac{\sigma(\zeta)}{e\Gamma+\sigma(\zeta)}\right)+\frac{k\ind{B}T\ln10}{e}(\text{pH}-\text{p}K\ind{a})-\frac{\sigma(\zeta)}{C}\label{eq:sig1}
\end{equation}
with the Boltzmann constant $k\ind{B}$ and the temperature $T=\SI{22}{\degree C}$, the surface density of chargeable sites $\Gamma=\SI{8}{\nm\tothe{-2}}$,\cite{Iler2,Hiemstra1989,Behrens2001} and the Stern layer capacitance $C=\SI{2.9}{\F\m\tothe{-2}}$.\cite{Hiemstra1989,Behrens2001} The terms $-\ln\left(\sigma/(e\Gamma+\sigma)\right)$ and $(\text{pH}-\text{p}K\ind{a})$ are only valid for negatively charged surfaces and are replaced by $\ln\left((e\Gamma-\sigma)/\sigma\right)$ and $(\text{p}K\ind{b}-\text{pH})$ if the surface charge becomes positive, i.e. below the isoelectric point pI at which the surface carries no net charge.\cite{Frament2013}

The surface charge density can be calculated using the planar Graham equation, if the electrical double layer (EDL) is not overlapping.\cite{Behrens2001,VanDerHeyden2005,Israelachvili2} This is valid if the Debye length is smaller than the pore tip radius, which is the case for every used electrolyte concentration except two, as shown in Table~\ref{t:debye}. The surface charge density is then given as
\begin{equation}
    \sigma(\zeta) = \frac{2\varepsilon\varepsilon_0}{\lambda\ind{D}}\frac{k\ind{B}T}{e}\sinh{\left(\frac{e\zeta(\sigma)}{2k\ind{B}T}\right)}\label{eq:sig2}
\end{equation}
with the permittivity of the liquid $\varepsilon=80.2$, the vacuum permittivity $\varepsilon_0$, and the Debye length $\lambda\ind{D}=\sqrt{\sum_i\varepsilon\varepsilon_0k\ind{B}T/e^2c_iz_i}$. $c_i$ and $z_i$ are the concentration and charge of the ionic species \ch{K+}, \ch{Cl-}, \ch{H+} and \ch{OH-}, respectively. Solving Equations~\ref{eq:sig1} and \ref{eq:sig2} self-consistently yields the $\zeta$-potential as well as the surface charge density as a function of the pH and the electrolyte concentration. Hence, in the nonlinear least-squares fitting of the conductance as a function of pH and concentration using Equation~\ref{eq:cond}, the $\text{p}K_a$, $\text{p}K_b$, and pI are fitting parameters and thus directly obtained. This is achieved by parallel fitting of the conductance as a function of pH and the conductance as a function of concentration (see below).

\begin{figure*}
 \centering
 \includegraphics[width=.99\textwidth]{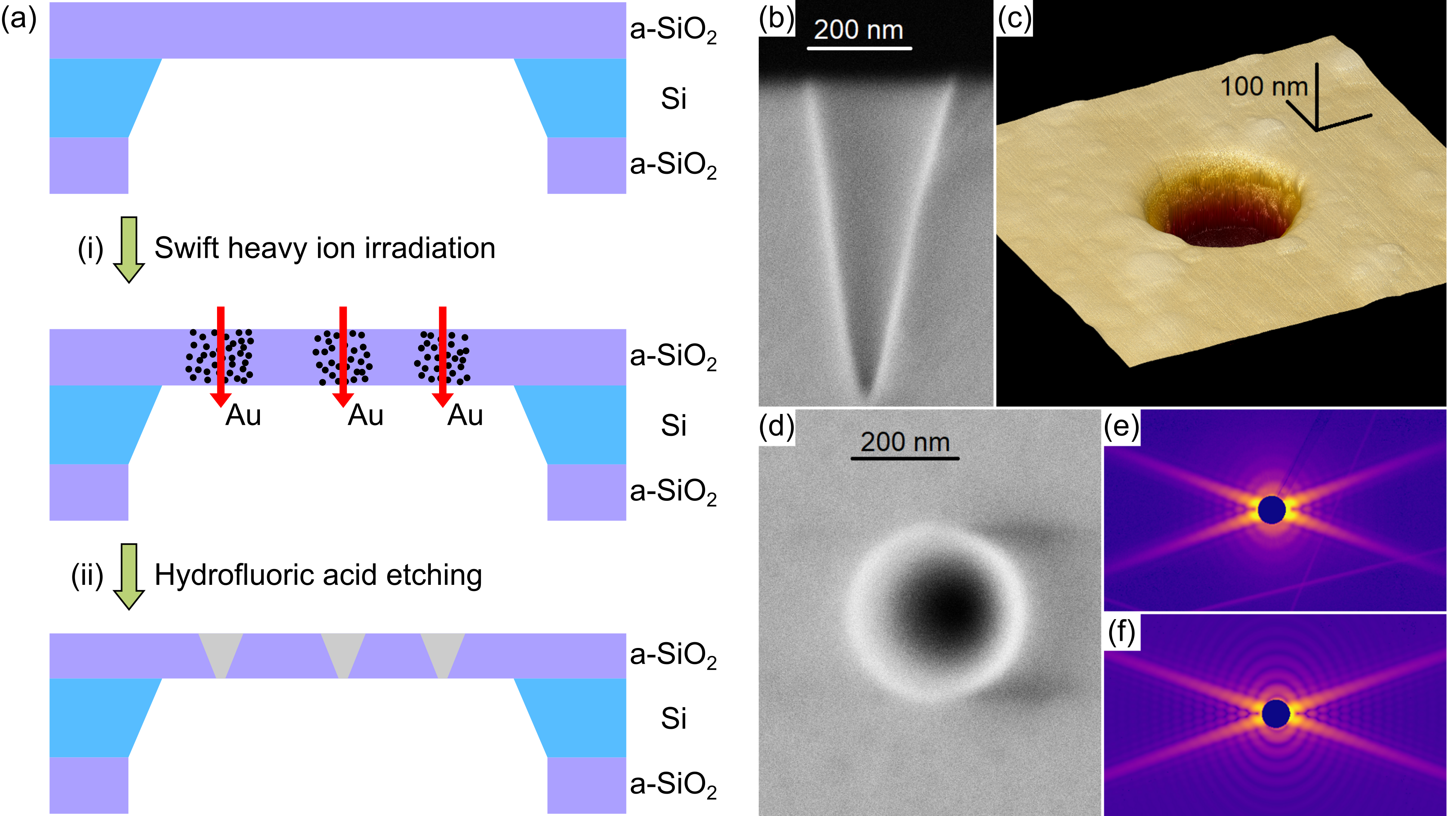}
 \caption{(a) Nanopore fabrication process using the track etch technique. (i) The a-\ch{SiO2} membrane is irradiated with swift heavy ions (Au ions), creating damage in the material. (ii) The damage is etched using hydrofluoric acid, forming conical nanopores. (b--f) Structural characterisation using scanning electron microscopy (SEM), atomic force microscopy (AFM), and small-angle X-ray scattering (SAXS). (b) SEM: cross section of a nanopore from a different membrane, fabricated under the same conditions. (c) AFM: high-resolution image of a nanopore. (d) SEM: plan view of a nanopore. (e,f) SAXS: (e) 2D scattering image and (f) fit to the data of a different membrane, fabricated under identical conditions.}
 \label{f:charact}
\end{figure*}

\section{Results and Discussion}
\subsection{Nanopore Fabrication}
The nanopores were created in \SI{920(5)}{\nm} thick, $55\times\SI{66}{\um\tothe{2}}$ a-\ch{SiO2} windows using the track etch technique,\cite{Hadley2018} which is schematically shown in Figure~\ref{f:charact}(a). First, the a-\ch{SiO2} windows were irradiated with \SI{2.2}{\GeV} \ch{^{197}Au} ions at the Universal Linear Accelerator UNILAC (GSI Helmholtz Centre for Heavy Ion Research Darmstadt, Germany) with a nominal fluence of \SI{1e6}{\text{ions }\cm\tothe{-2}}. Because of the small membrane size, fluctuations in the number of pores occur, and the real fluence across the membrane was \SI{4.4e5}{\text{ions }\cm\tothe{-2}} based on the number of pores found (i). When passing through the material, the ions generate a narrow cylindrical damage region along their path.\cite{Kluth2008} These so-called ion tracks are more susceptible to chemical etching than the undamaged material. After irradiation, the membrane was taped to a silicon wafer using Kapton tape and immersed in \SI{2.5}{\percent} hydrofluoric acid at room temperature, which removes material along the track thus at a higher rate than the surrounding matrix (ii).\cite{Hadley2020} The setup used for etching is described in more detail in the SI (Figure~S1). This process creates highly uniform conical nanopores in the thin a-\ch{SiO2} windows. To stop the etching, the membranes were rinsed with and stored in deionised water until they were characterised.

\begin{figure*}
 \centering
 \includegraphics{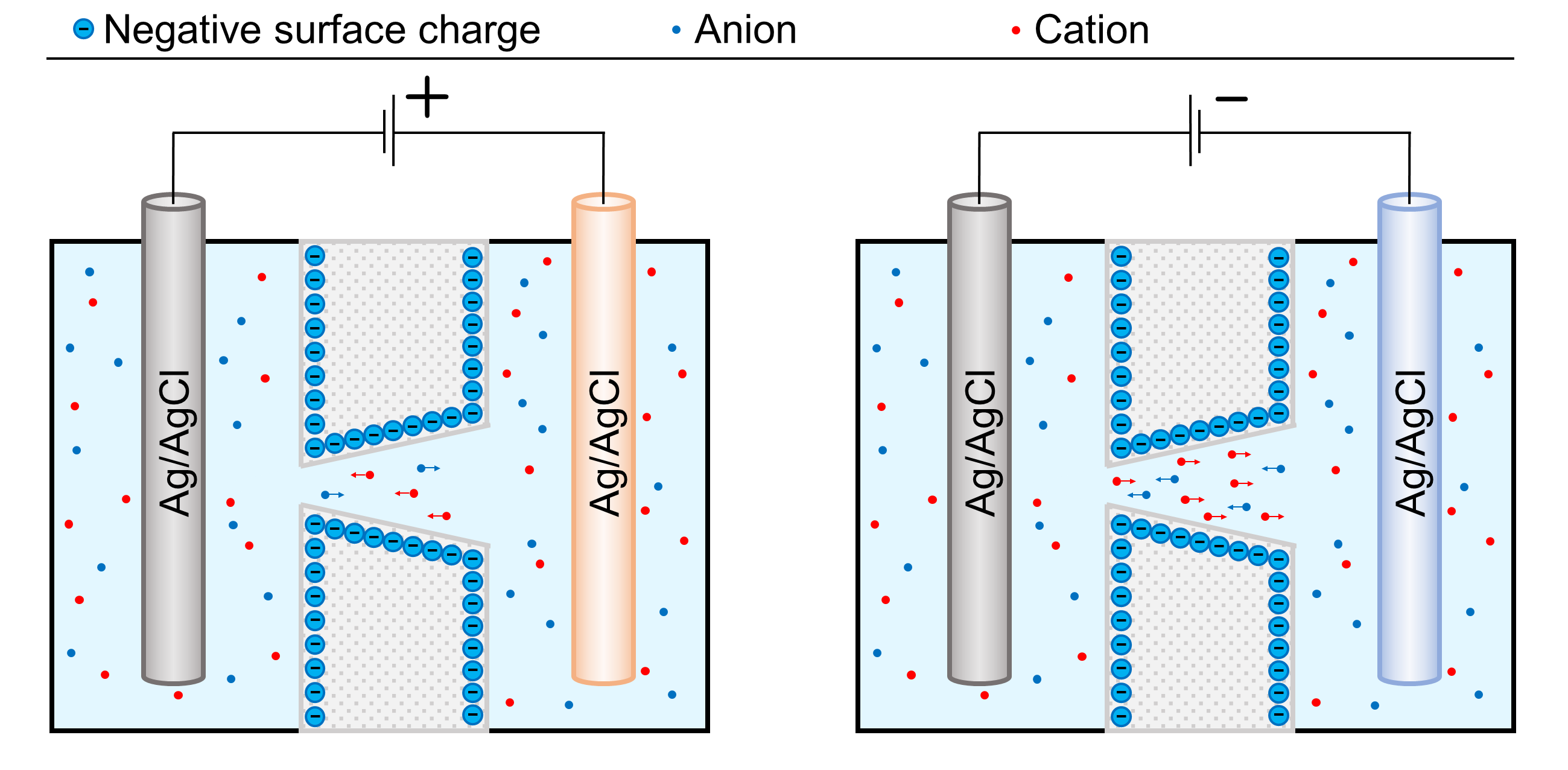}
 \caption{Conductometric nanopore experiments: Scheme of the setup with a negatively charged nanopore membrane. Because of the negative surface charge, there are more cations than anions transported through the pore. In the positive case (left), the pore is ion depleted, representing the low conductance state. For negative voltages (right), the concentration inside the pore is increased, corresponding to the high conductance state.}
 \label{f:setup}
\end{figure*}

With this fabrication technique, we can manufacture nanopore membranes with a wide range of different morphologies. By varying the thickness of the a-\ch{SiO2} layer, nanopores with lengths between 200 and \SI{1000}{\nm} can be created. By combining a variable irradiation fluence of up to \SI{e9}{\text{ions }\cm\tothe{-2}} and window sizes ranging from $30\times30$ to $700\times\SI{700}{\um\tothe{2}}$, membranes containing between 10 and 5,000,000 nanopores can potentially be achieved. Single ion irradiation leading to a single pore is also possible.\cite{Chtanko2004} The half-cone angle can be tuned between \SI{10}{\degree} and \SI{18}{\degree} by using different irradiation energies and adjusting the etching conditions.\cite{Hadley2019} The nanopore etching process is very reproducible, as demonstrated previously for etch pits in thin films of a-\ch{SiO2}.\cite{Hadley2018,Hadley2019,Hadley2020} This technique has now been translated to a-\ch{SiO2} membranes. The biggest variance between membranes is the number of pores for low irradiation fluences. While aiming for a fluence of \SI{1e6}{\text{ions }\cm\tothe{-2}}, which would yield 36 pores, we obtained only 16 pores. Analysing multiple membranes reveals that up to 50 pores per membrane can be created at this nominal fluence. This statistical variance is negligible at fluences above \SI{1e7}{\text{ions }\cm\tothe{-2}}.

Here, we are presenting a detailed structural characterisation and the ionic transport properties of conical nanopores in an a-\ch{SiO2} membrane. Further descriptions of the experiments are given in the SI.

\subsection{Structural Characterisation}
Scanning electron microscopy (SEM) is a useful tool to estimate the pore base radius, number of pores, and pore distribution across the nanopore membrane. Figures~\ref{f:charact}(b) and \ref{f:charact}(d) show a cross section and a plan view of a conical nanopore. To acquire a cross-section, the membrane is broken in half. As this is a destructive technique, Figure~\ref{f:charact}(b) is from a different membrane fabricated under identical conditions. From the SEM data, we conclude that the membrane contains 16 nanopores with an average minimum distance to the neighbouring pore of \SI{7.8(46)}{\um} (see SI, Figure~S2) and a pore base radius of \SI{169(7)}{\nm} (see SI, Figure~S3). The uncertainties are the standard deviations obtained from averaging over multiple pores. The spacing of the nanopores is sufficient to rule out any meaningful contributions of interpore interactions.\cite{Cao2018}

The base radius of the pores can also be determined using atomic force microscopy (AFM). As presented in Figure~\ref{f:charact}(c), AFM provides a high-resolution image of the nanopore, from which a base radius of \SI{158.3(26)}{\nm} is obtained. The uncertainty is the standard deviation obtained by averaging over multiple pores. The difference in base radius compared to the SEM measurements might be due to the difficulty in defining the edge of the pore in either measurement technique and limited resolution. For more information, see the SI (Figures~S4 and S5).

Using ellipsometry (JA Woollam M-200D ellipsometer) and surface profiling (Bruker Dektak surface profiler), the membrane thickness and hence the pore length were determined to \SI{710(5)}{\nm}, indicating a thinning down from the original thickness of \SI{920(5)}{\nm} by \SI{210(5)}{\nm} during the etching process. With an etching time of \SI{14.5}{\min}, this results in a bulk etch rate of \SI{14.5(4)}{\nm\min\tothe{-1}}. Based on the pore geometry and the kinetics of the etching process, this allows the computation of the track and radial etch rate to \SI{66.5(19)}{} and \SI{11.6(3)}{\nm\min\tothe{-1}}, respectively.\cite{NIKEZIC2004,Dutt2023} Due to the interference of the ellipsometry laser with the nanopores, the thickness of the membrane could not determined directly by ellipsometry only. Thus, the step between an unetched area of the membrane was compared to the etched area using surface profiling.

To gain precise measurements of the cone angle of the nanopores, synchrotron-based small-angle X-ray scattering (SAXS) was performed at the SAXS/WAXS beamline at the Australian Synchrotron (Melbourne, Australia) with an X-ray energy of \SI{12}{\keV}. By tilting the membrane with respect to the X-ray beam and analysing the change in scattering pattern, the cone angle of the pores can be determined. This measurement has been performed using a different membrane that was fabricated under identical conditions irradiated with a fluence of \SI{e8}{\text{ions }\cm\tothe{-2}} to obtain a sufficiently high SAXS signal. It can be assumed that the etching process and the SAXS measurements are unaffected by the difference in fluence, as in both cases the pore overlap is negligible.\cite{Kluth2008b} Figure~\ref{f:charact}(e) shows the 2D scattering image with the characteristic x-wing pattern and oscillations in intensity for a tilt angle of \SI{41}{\degree}. By performing a fit of the 2D scattering data (Figure~\ref{f:charact}(f)) to our theoretical model, we find that the half-cone angle of the nanopores is \SI{12.6(1)}{\degree} (more SAXS measurements are presented in the SI, Figure~S6). A detailed explanation of the measurement and data analysis technique is given by Hadley et al.\cite{Hadley2018}

Utilising the structural characterisation using the different measurement techniques, we can predict the tip radius of the nanopore $r\ind{t}$ using the relation
\begin{equation}
    r\ind{t}=r\ind{b}-L\tan\vartheta
\end{equation}
where $r\ind{b}$ is the base radius, $L$ is the length, and $\vartheta$ is the half-cone angle of the nanopore. The tip radius calculated by this method varies between 0 and \SI{11}{\nm} due to the uncertainties determining the base radius using AFM and SEM (a more comprehensive analysis is given in the SI, Figure~S7). It is important to note that the cone angle determined by SAXS has an uncertainty of only \SI{.1}{\degree} and thus does not contribute significantly to this. Conductometry measurements, however, will provide a better measure for the tip radius, which is presented in the following section.

\begin{figure}
    \begin{minipage}{.5\textwidth}
        \includegraphics[width=.99\textwidth]{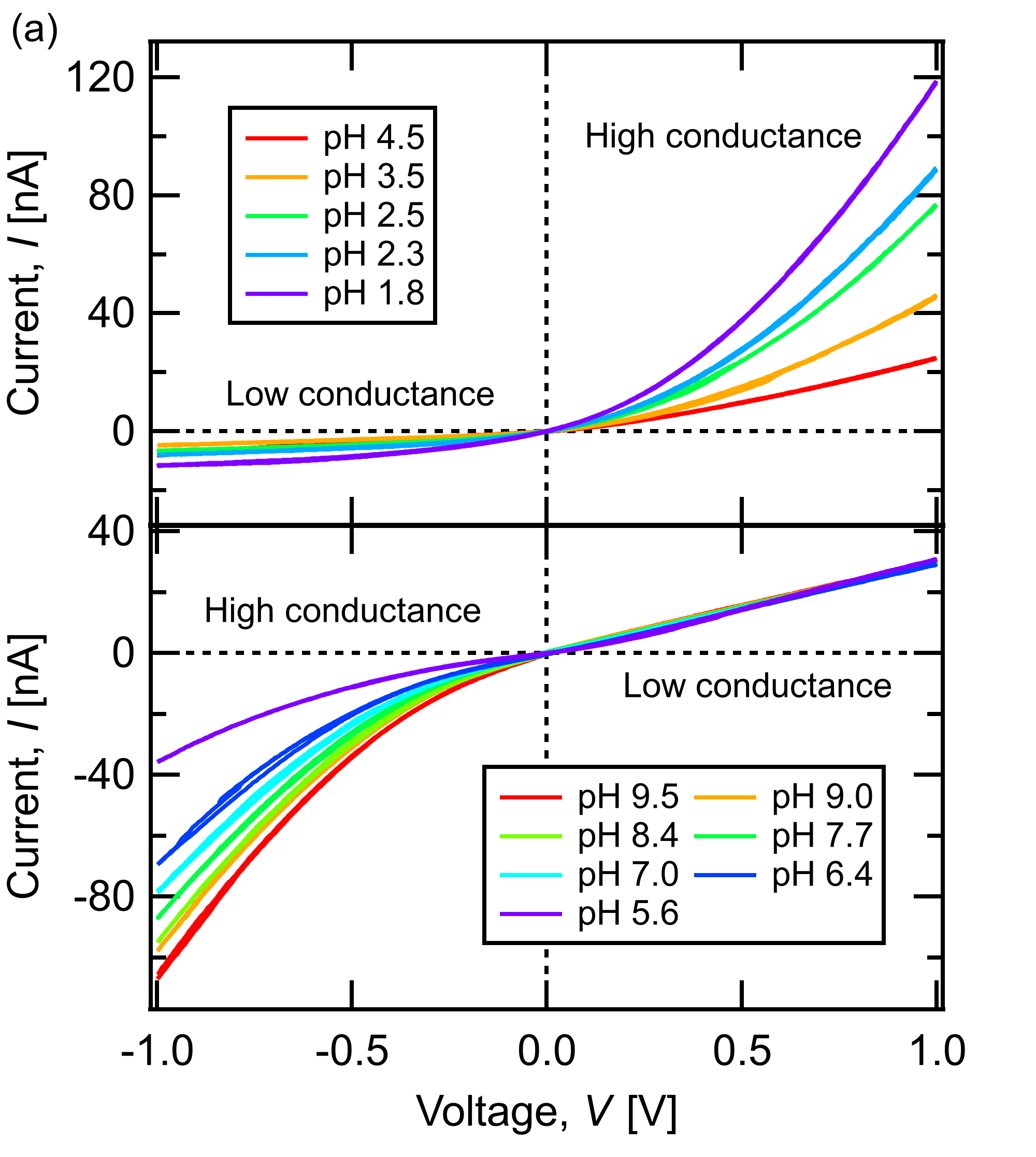}
    \end{minipage}
    \begin{minipage}{.5\textwidth}
        \includegraphics[width=.99\textwidth]{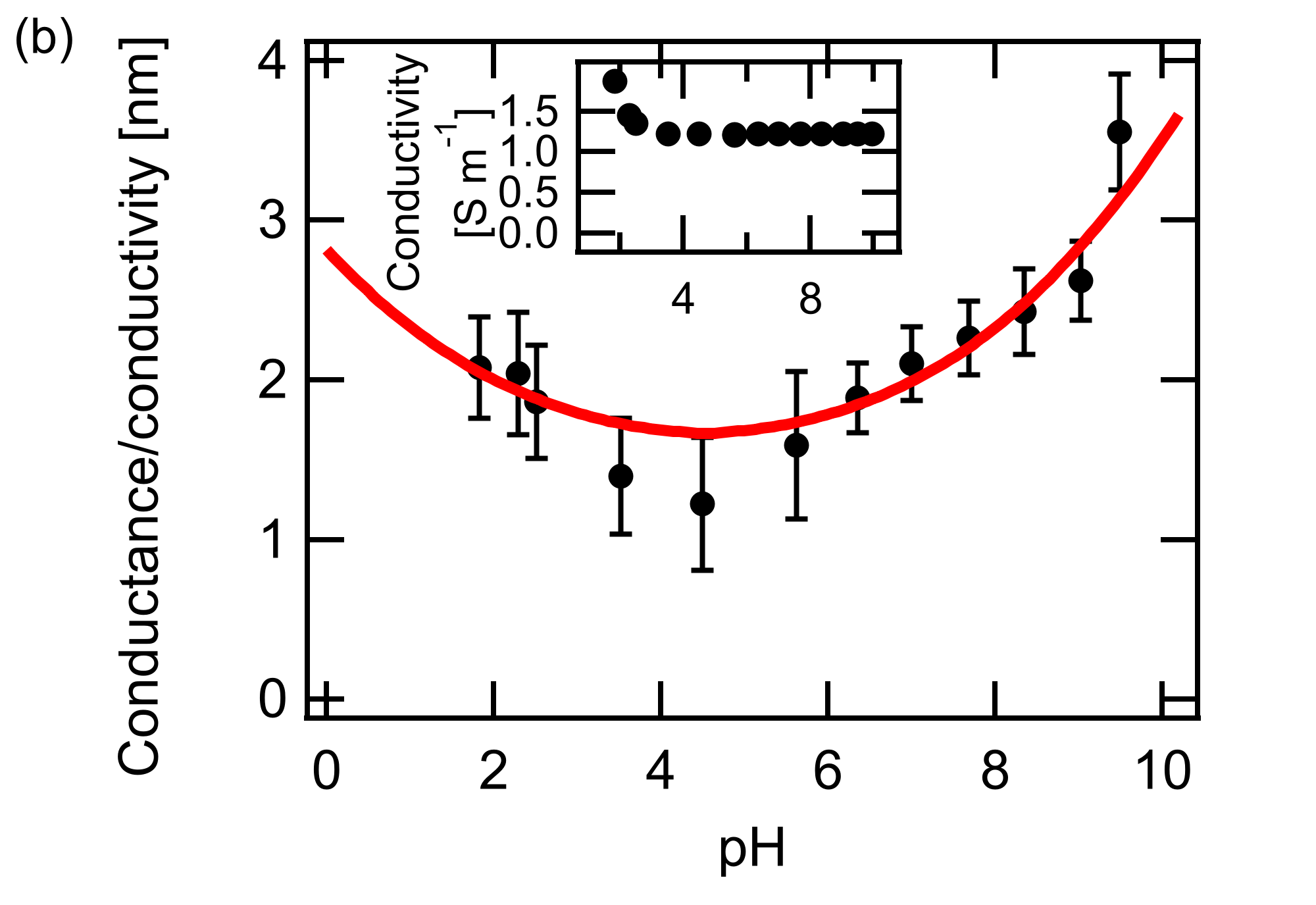}
        \includegraphics[width=.99\textwidth]{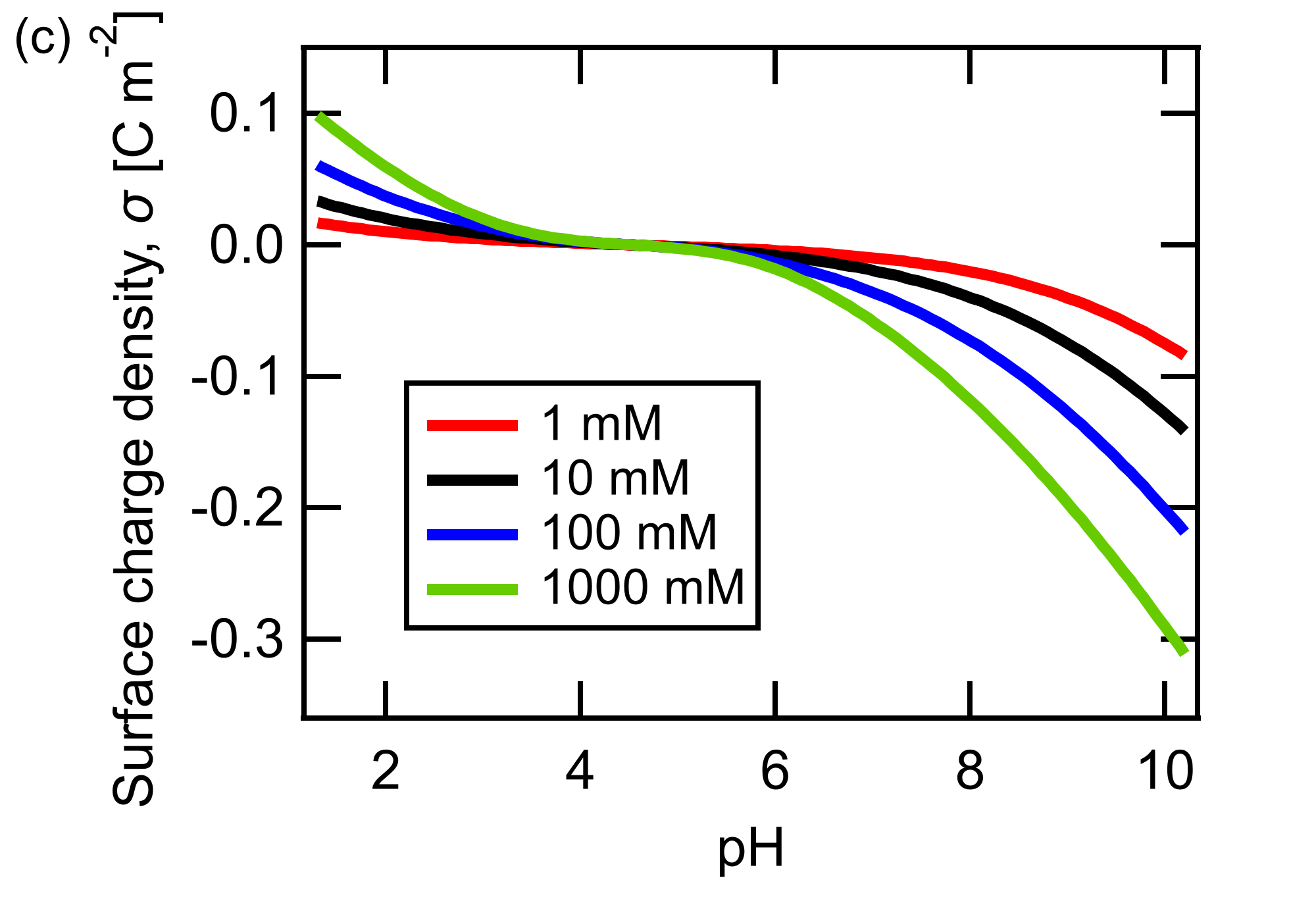}
    \end{minipage}
  \caption{(a) $I-V$ curves in \SI{100}{\mymmol} \ch{KCl} for the pH series in acidic and basic conditions. (b) Experimental membrane conductance (symbols) including fit to Equation~\ref{eq:cond} (lines) for the pH series at a concentration of \SI{100}{\mymmol}. The conductance is normalised by the electrolyte conductivity, as the conductivity slightly varies with pH (inset). (c) Calculated surface charge densities as a function of pH for different concentrations for our nanopore system. The surface charge densities were calculated using the experimentally obtained $\text{p}K_a$, $\text{p}K_b$, and pI values and Equations~\ref{eq:sig1} and \ref{eq:sig2}.}
  \label{f:ph}
\end{figure}

\subsection{Ion Transport and Surface Properties}
The ionic transport characteristics of the nanopores were analysed as a function of electrolyte concentration and pH conditions. The measurements were conducted by inserting the nanopore membrane between two compartments, which are filled with a potassium chloride (\ch{KCl}) solution at concentrations ranging from \SIrange{.1}{1000}{\mymmol} and pH values between \numlist{1.8;10.1}{}. The pH was adjusted by adding \ch{HCl} or \ch{KOH}. At room temperature, the etch rate of \ch{SiO2} at these pH values is $\ll$\,\SI{1}{\nm\,\text{h}\tothe{-1}}, and thus no impact on the size or shape of the pores is expected.\cite{Chung2007,koh} Using \ch{Ag}/\ch{AgCl} electrodes on either side of the membrane, we measured the current--voltage ($I-V$) characteristics of the pores between \SIlist{-1;1}{\V} to determine the membrane conductance and ion current rectification (ICR). The setup is described in more detail in the SI (Figure~S8). The conductance was determined in the linear regime of the $I-V$ curve around \SI{0}{\V} between $-$0.1 and \SI{.1}{\V}. The convention for conical nanopore experiments is that the polarity of the voltage is determined by the electrode at the cone base, as shown in Figure~\ref{f:setup}.

In a charged nanopore, the concentration of counterions to the surface charge is greatly increased, and thus the majority of the ionic current is carried by them. As the electric field in the tip region of a conical nanopore is greatly increased, transport of majority charge carriers from tip to base is promoted, and a large ionic current is observed. This case is called the high conductance state. If the polarity of the electric field is reversed, transport is hindered, which is termed the low conductance state.

Figure~\ref{f:ph}(a) shows the $I-V$ curves for the pH series. A magnified plot depicting the current around \SI{0}{\V}, which was used to determine the membrane conductance, is presented in the SI (Figure~S9). In acidic conditions, the high conductance state of the membrane occurs at positive voltages, indicating a positive surface charge which is the opposite in basic conditions, indicating a negative surface charge. In either case, the high conductance state is strongly dependent on the pH, while the low conductance state is almost unaffected. In the high conductance state, the ionic current is dominated by surface effects, indicating a large change in surface charge with varying pH. As in the low conductance state the current is dominated by bulk behaviour, the ionic current is almost unaffected by changes in pH.

Figure~\ref{f:ph}(b) shows the membrane conductance as a function of pH. The symbols are our measurements, and the solid line represents a fit to Equation~\ref{eq:cond}. The membrane conductance is normalised by the electrolyte conductivity as it changes slightly with pH (inset). The minimum conductance occurs at a pH of 4.5, at which the surface charge density reaches 0\,C\,m$^{-2}$ as plotted in Figure~\ref{f:ph}(c). Figure~\ref{f:ph}(c) shows the surface charge densities that are calculated using the experimentally obtained $\text{p}K_a$, $\text{p}K_b$, and pI values and Equations~\ref{eq:sig1} and \ref{eq:sig2}. At the point of zero surface charge density, which is the isoelectric point pI, the pores only exhibit bulk current behaviour. In more acidic conditions, the surface becomes positively charged, and in more basic conditions, the pore becomes negatively charged.

\begin{figure}
    \begin{minipage}{.5\textwidth}
        \includegraphics[width=.99\textwidth]{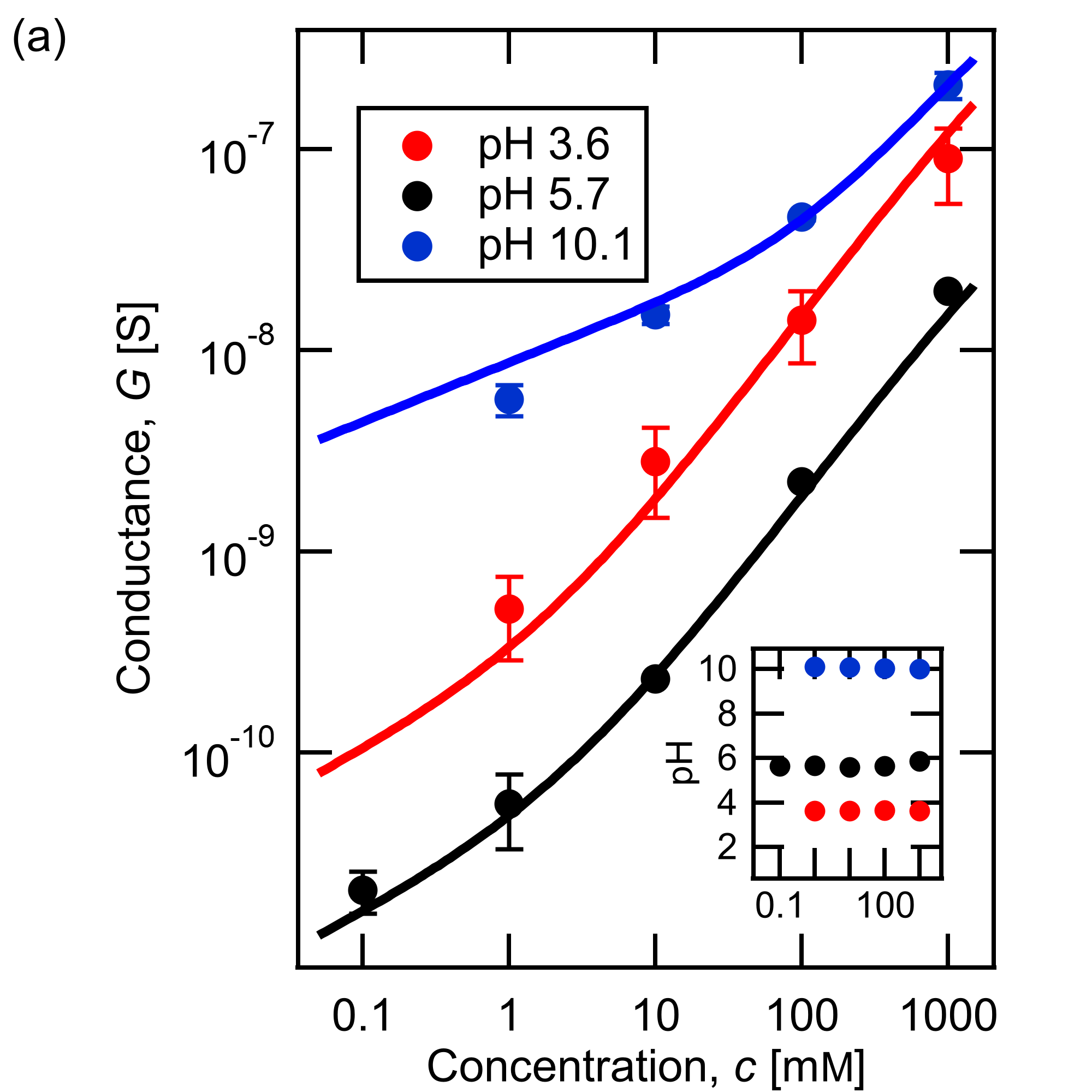}
    \end{minipage}
    \begin{minipage}{.5\textwidth}
        \includegraphics[width=.99\textwidth]{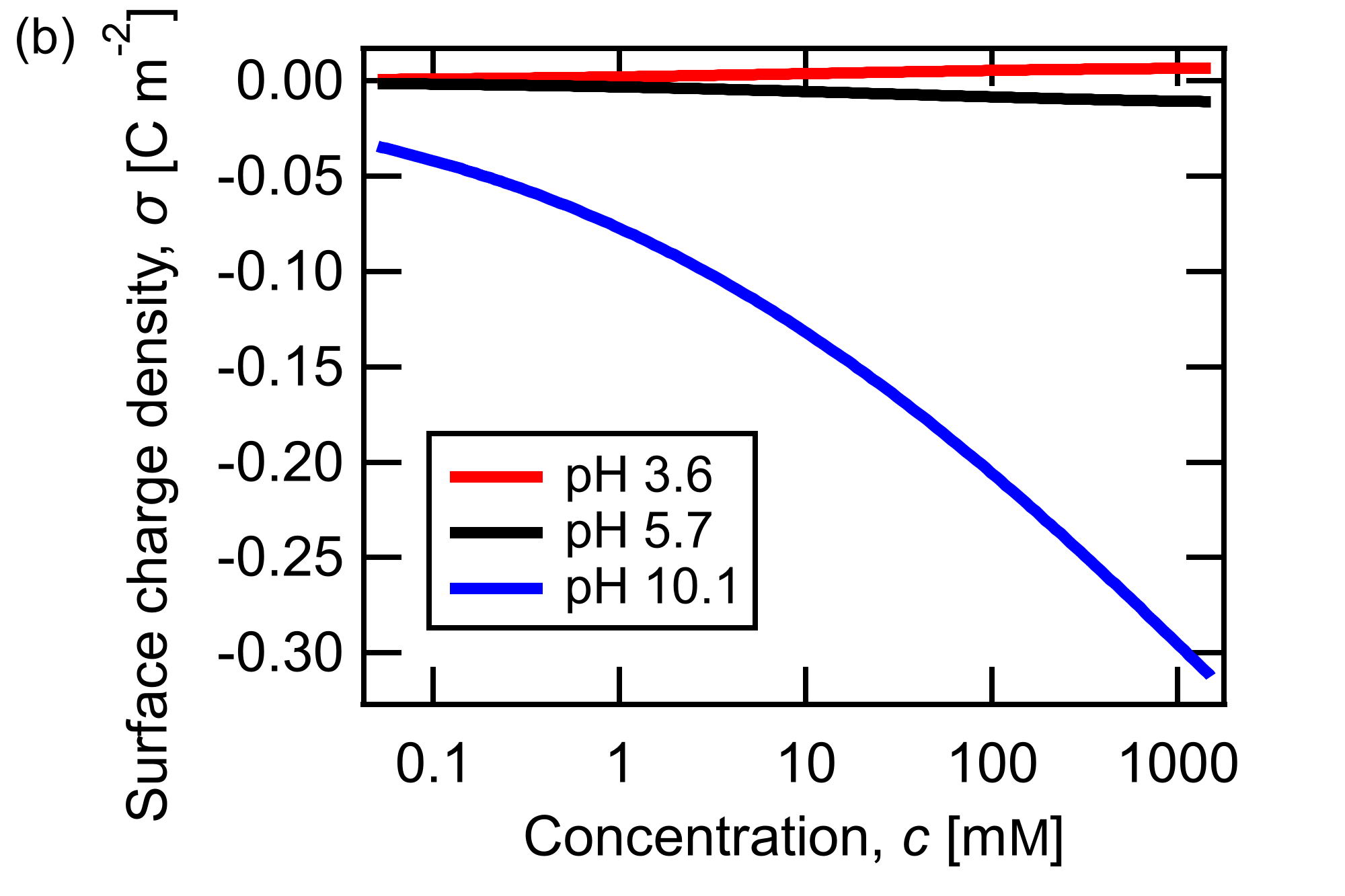}
    \end{minipage}  
  \caption{(a) Experimental membrane conductance (symbols) including fit to Equation~\ref{eq:cond} (lines) for the concentration series for three different pH values, and pH as a function of concentration (inset). (b) Calculated surface charge densities as a function of concentration for different pH values for our nanopore system. The surface charge densities were calculated using the experimentally obtained $\text{p}K_a$, $\text{p}K_b$, and pI values and Equations~\ref{eq:sig1} and \ref{eq:sig2}.}
  \label{f:concentration}
\end{figure}

\begin{table}
  \caption{Debye length $\lambda\ind{D}$ (calculated using Equation~\ref{eq:sig2}) for different KCl concentrations. Extension of the EDL through the nanopore at the pore tip indicated by the ratio of Debye length to tip radius $\lambda\ind{D}/r\ind{t}$ for a tip radius of \SI{5.7}{\nm}.}
  \label{t:debye}
  \begin{tabular}{lll}
    \hline
    concentration, $c$ [\SI{}{\mymmol}]  & Debye length, $\lambda\ind{D}$ [\SI{}{\nm}]  & $\lambda\ind{D}/r\ind{t}$ [$\%$]  \\
    \hline
    0.1 & 30.8 & 540 \\
    1 & 9.72 & 171 \\
    10 & 3.07 & 54 \\
    100 & 0.97 & 17 \\
    1000 & 0.31 & 5.4 \\
    \hline
  \end{tabular}
\end{table}

Figure~\ref{f:concentration}(a) shows the membrane conductance as a function of electrolyte concentration for three pH values. The symbols are our measurements, and the solid line is the fit to Equation~\ref{eq:cond}. We note that due to the contribution of \ch{HCl} and \ch{KOH} when adjusting the pH, for pH values of 3.6 and 10.1, the concentration of \ch{KCl} was limited to \SI{1}{\mymmol}. At \SI{1}{\mymmol} adjusting the pH changes the concentration of \ch{Cl-} and \ch{K+} by approximately \SI{25}{\percent}. However, this does not alter the results and is included in the uncertainty calculations (see below). At a pH of 3.6, the surface is positively charged, at a pH of 5.7, the pores are slightly negatively charged, and at a pH of 10.1, the pores are strongly negatively charged, which is reflected in Figure~\ref{f:concentration}(b). At high concentrations, the overall conductance is dominated by bulk behaviour, as the Debye lengths becomes smaller (see Table~\ref{t:debye}). The Debye length describes the screening length of the surface charge; a reduced Debye length is correlated to a thinner electrical double layer (EDL). At lower concentrations, where the slope of the conductance changes, surface effects start to dominate over bulk behaviour.

By parallel analysis and nonlinear least-squares fitting of the pH- and concentration-dependent membrane conductance with Equation~\ref{eq:cond}, the $\text{p}K_a$, $\text{p}K_b$, and pI values as well as pore size can be determined. During fitting, the data were weighted by the standard deviation, which was obtained by repeating each measurement twice. We obtain a pore size of \SI{5.7(1)}{\nm}, where the uncertainty is the \SI{95}{\percent} confidence interval of the fit. This is well within the range of the pore size determined by the structural characterisation. The $\text{p}K_a$, $\text{p}K_b$ and pI are \num{7.6(1)}, \num{1.5(2)}, and \num{4.5(1)}, respectively, demonstrating that nanopores in track etched a-\ch{SiO2} exhibit similar properties to a-\ch{SiO2} nanoparticles and nanochannels, for which $\text{p}K_a$, $\text{p}K_b$, and pI values of 6.6~--~8, 0~--~2 and <~5, respectively, have been reported.\cite{VanDerHeyden2005,Wang2010,Andersen2011,Yeh2013,Barisik2014} The actual pI value is most likely slightly above 4.5, as the $I-V$ curve at a pH of 4.5 is still slightly nonlinear, indicating a positively charged surface. As the surface charge is extremely sensitive to the pH around the pI, even a slight deviation will cause a surface charge that renders the $I-V$ curve nonlinear. This deviation is reflected in the uncertainty of our reported pI value.

\subsection{Ionic Current Rectification}
Ionic current rectification (ICR) is an important characteristic of (asymmetric) nanopores in which the membrane exhibits a directional preference for transport of charged species.\cite{Siwy2006,Siwy2007,Vlassiouk2008,Gamble2014} ICR can be achieved by different means, such as by a varying surface charge density along the nanopore,\cite{Daiguji2005} asymmetries in the electrolyte,\cite{Hsu2018} or geometry of the pores such as our conical nanopores. ICR is defined as the ratio between the current in the high conductance state to the low conductance state:
\begin{equation}
    \text{ICR} = 
    \begin{cases}
    \left|\frac{I\left(\SI[retain-explicit-plus]{+1}{\V}\right)}{I\left(\SI{-1}{\V}\right)}\right|&\text{if pH}<\text{pI}\\
    \left|\frac{I\left(\SI{-1}{\V}\right)}{I\left(\SI[retain-explicit-plus]{+1}{\V}\right)}\right|&\text{if pH}>\text{pI}
    \end{cases}
\end{equation}

\begin{figure}
    \begin{minipage}{.5\textwidth}
        \includegraphics[width=.99\textwidth]{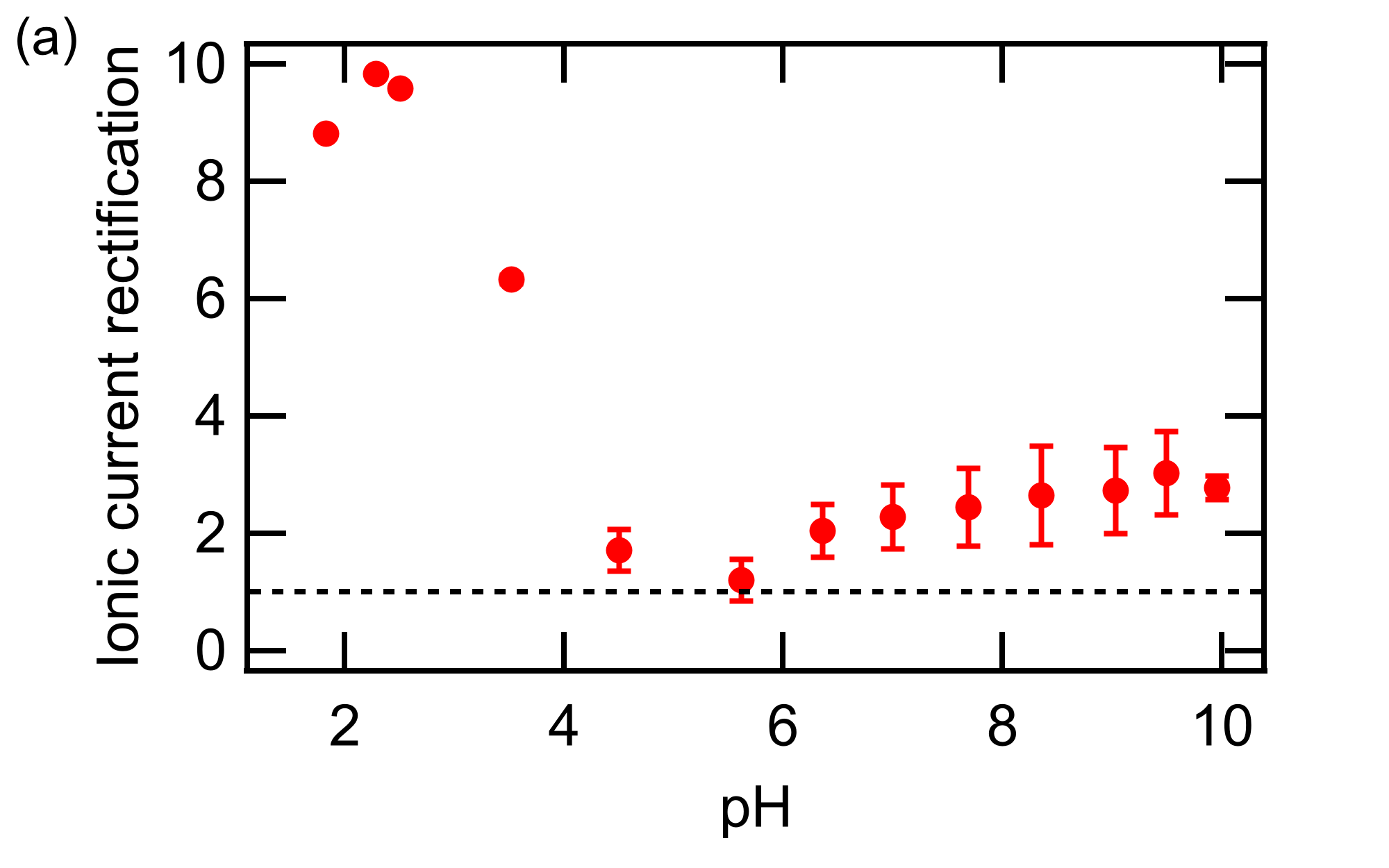}
    \end{minipage}
    \begin{minipage}{.5\textwidth}
        \includegraphics[width=.99\textwidth]{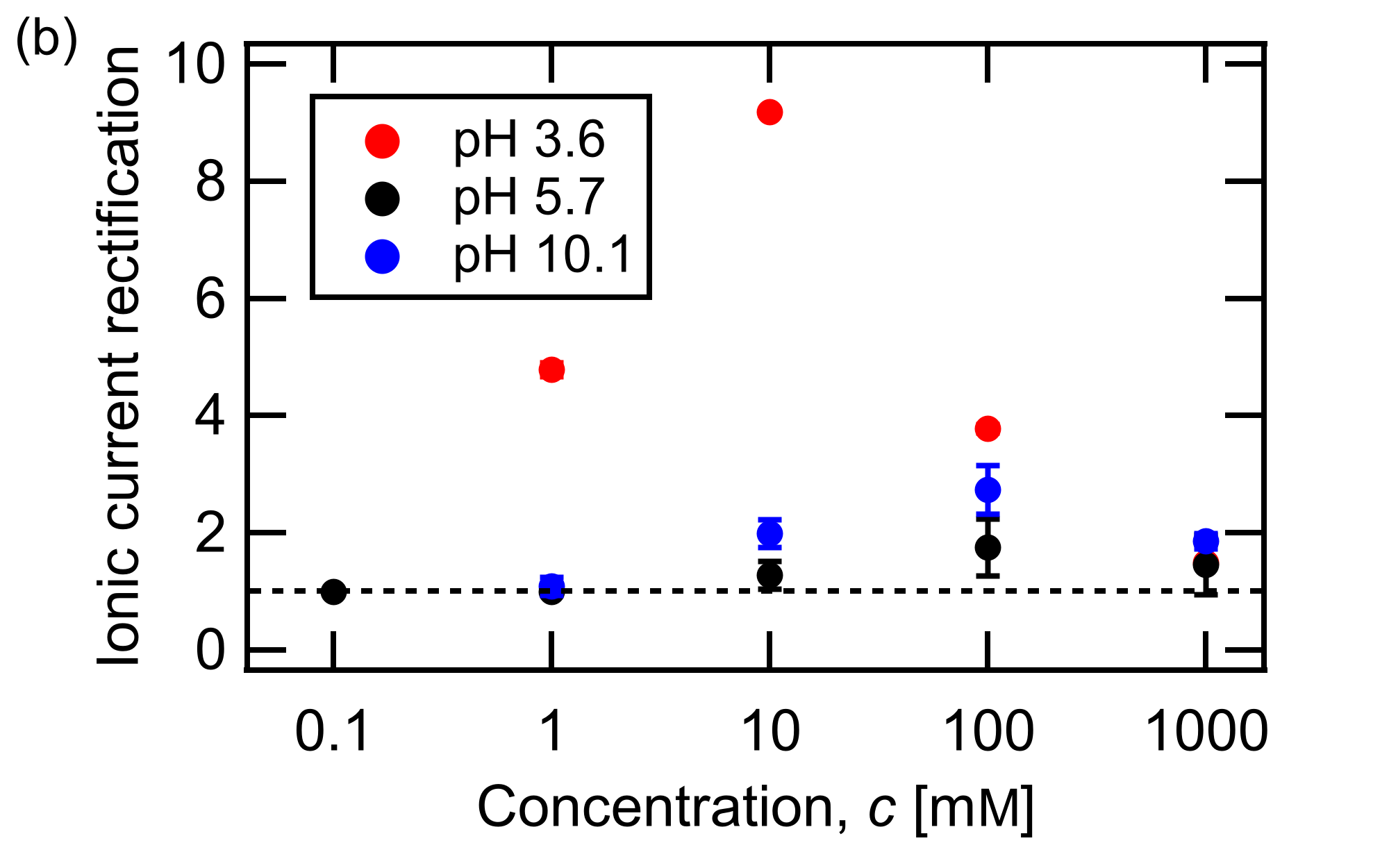}
    \end{minipage}
  \caption{Ionic current rectification (ICR) as a function of (a) pH at a concentration of \SI{100}{\mymmol} and (b) concentration for three different pH values. The dashed line indicates a rectification ratio of 1, i.e., no rectification and a symmetric current.}
  \label{f:icr}
\end{figure}

Figure~\ref{f:icr} shows the ICR of our membrane as a function of pH and concentration. The rectification ratio peaks at around 10 in acidic conditions at a pH between 2 and 3 at concentrations between 10 and \SI{100}{\mymmol}, corresponding to surface charge densities of 0.03~--~\SI{.04}{\coulomb\m\tothe{-2}}. This behaviour was theoretically predicted by Tseng et al.\cite{Tseng2016} In the high conductance state, when the surface charge is low, an increase in surface charge will cause a large increase in the concentration of ions within the pore and thus the current. Conversely, in the low conductance state, the concentration of ions barely increases by increasing the surface charge, and thus the current remains almost constant. This effect can be observed in our measurement. Figure~\ref{f:ph}(a) shows a large change in the high conductance state between pH values of 4.5 and 2.3, whereas the low conductance state is almost stable and thus ICR increases. However, when the surface charge is increased further, the concentration and thus current saturates in the high conductance state. In the low conductance state, the concentration and thus current will increase significantly. We observe this effect when adjusting the pH from 2.3 to 1.8, and hence ICR decreases. Thus, for any given pore geometry, ICR will have a maximum value at a certain surface charge density, which is 0.03~--~\SI{.04}{\coulomb\m\tothe{-2}} in our nanopore system.

\subsection{Comparison to Other Nanopore Systems}
The conical nanopores in a-\ch{SiO2} presented here have the potential to show increased performance compared to other asymmetric nanopore systems. As polymer membranes are the most commonly used asymmetric multipore membranes, they provide a good baseline for comparison. Based on the geometry of our nanopores, it is possible to compare the conductance and ICR as well as the potential performance of different applications such as electroosmotic pumps and nanofluidic osmotic power generation. To do this, it will be assumed that tip size and surface charge will be the same, as similar values have been reported in other systems.\cite{Laucirica2021} The only differences stem from the cone angle and pore length.

 The pore conductance is inversely proportional to the square root of the pore length $G\propto\sqrt{1/L}$ and proportional to the square root of the half-cone angle $G\propto\sqrt{\vartheta}$. Our \SI{710}{\nm} pore should have a conductance that is higher by a factor of 1.7, 3.8, and 6.5 than a polymeric pore of lengths 2, 10, and \SI{30}{\um}, respectively. Similarly, a \SI{12.6}{\degree} pore should have a conductance that is higher by a factor of 3.5, 2.0, and 1.6 than a polymeric pore of cone angles \SI{1}{\degree}, \SI{3}{\degree}, and \SI{5}{\degree}, respectively. Unfortunately, a direct comparison to literature values is not possible as this would require a pore with exactly the same size and surface charge and the experiment to be conducted under precisely the same conditions. Combining both a reduced length and higher cone angle increases the conductance even further. Additionally, if the experimental conditions are optimised, ideal ICR occurs at half-cone angles between \SI{10}{\degree} and \SI{15}{\degree}.\cite{Tseng2016} Thus, a-\ch{SiO2} pores should perform better than pores in polymers in both aspects.

 It has been predicted that the ideal pore length for power generation is in the range of 400~--~\SI{1000}{\nm}.\cite{Laucirica2021} Above that, the membrane resistance starts to greatly reduce the power output. For thinner membranes, concentration polarisation drastically impacts the performance.\cite{Laucirica2021} Our nanopores reside within this optimal length window. Another problem for making nanofluidic osmotic power generation viable is scaling the performance of a single nanopore to a larger number of pores. With increasing areal pore density, pores get closer to each other. At a certain distance, interpore effects greatly decrease the overall performance. It has been shown that an interpore distance of \SI{500}{\nm} yields the maximum power output.\cite{Laucirica2021} With a pore base radius of \SI{164.4}{\nm}, the average minimum distance to the closest neighbour is achieved at a fluence of \SI{3.7e7}{\text{ions }\cm\tothe{-2}}. The polymer membrane that was used by Guo et al. for power generation had a thickness of \SI{12}{\um}, a half-cone angle of \SI{2.8}{\degree}, and a base radius of \SI{600}{\nm}.\cite{Guo2010} To achieve an ideal interpore distance, only a fluence of \SI{.9e7}{\text{ions }\cm\tothe{-2}} is possible. Thus, with a 4 times higher pore density as well as a reduced length, our a-\ch{SiO2} system offers a potential gain in performance. A disadvantage of our membrane platform compared to polymers is the limited size of our membranes. Above $700\times\SI{700}{\um\tothe{2}}$, the membranes become too fragile to work with. The size of polymer membranes is only limited by the capabilities of the ion accelerator facility, and membranes with diameters of many centimetres have been fabricated. However, a-\ch{SiO2} membranes are ideal for smaller systems and can be readily integrated into lab-on-a-chip devices. 

 As the pressure driven flow $Q$ of an electroosmotic pump is proportional to the cone angle $Q\propto\tan{\vartheta}$, it can be estimated that our system would provide a 7.1 times higher flow than a similar system in a polymer with a cone angle of \SI{1.8}{\degree}.\cite{Wu2016} Additionally, the same argument regarding scalability as for power generation can be made. Because of the larger possible fluence without causing too much interpore interaction, a higher pore density and thus performance can be achieved.

 Another technique with which conical nanopores can be fabricated is electron beam lithography (EBL) followed by reactive ion etching (RIE) and chemical etching.\cite{Zeng2019a,Chuah2019} The big advantage of this technique is the ability to precisely determine the location of each individual nanopore. Thus, ideal placement of nanopores can be guaranteed, i.e., equidistant spacing and distribution across the membrane. This is not the case for track etched membranes, where the pores are randomly distributed. However, this fabrication method comes with its own drawbacks. The limited speed of EBL makes large-scale fabrication unfeasible. Exposing an area of just \SI{1}{\cm\tothe{2}} can take up to 12 days.\cite{Parker2000} This severely limits the total number of pores that can be reasonably fabricated. Another issue is the lack of customisability of the nanopores. Being able to adjust the cone angle of the pores is crucial, as depending on the desired application and other factors such as geometrical properties or electrolyte conditions, different cone angles are required for optimisation. An example is \ch{SiN}, where the cone angle cannot be readily changed.\cite{Wei2010} In contrast, we can tune the cone angle of our nanopores using different ion irradiation energies, ion species, or different HF concentrations for etching.\cite{Hadley2019}

\section{Conclusions}
We presented a detailed characterisation of conical nanopores in a-\ch{SiO2} that were fabricated using the track etch technique. The geometrical structure of the nanopore membrane was characterised using SAXS, AFM, SEM, ellipsometry, and surface profiling. The multipore membrane used in this research contains 16 nanopores with tip radii of \SI{5.7(1)}{\nm}. We determined the important surface parameters $\text{p}K_a$, $\text{p}K_b$, and pI, which are \num{7.6(1)}, \num{1.5(2)}, and \num{4.5(1)}, respectively, which are in good agreement with those from other \ch{SiO2} systems. The maximum ICR was determined to be around 10 at pH 2.3. 

It is important to note that these measurements have been performed with a multipore system. Frequently, characterisations are performed using single nanopores, and scaling those to multipores can lead to a significant loss in performance.\cite{Cao2018,Wang2021} ICR is an extremely important property of nanopore systems, as ICR enables the development of innovative applications as well as improving the performance of nonrectifying membranes. ICR has been utilised to create new biosensors\cite{Cai2015a,Perez-Mitta2018} as well as demonstrate superior performance in power generation.\cite{Li2019,Laucirica2021} Electroosmotic pumps (EOPs) that operate in the AC mode have also been developed.\cite{Wu2016} We have shown that our system has the potential to show performance gains compared to existing nanopore platforms in terms of conductance, ICR, EOP performance, and nanofluidic osmotic power generation. Additionally, a-\ch{SiO2} membranes could be scaled to larger pore densities compared to some polymers without compromising performance compared to different materials.

a-\ch{SiO2} membranes can be readily integrated into lab-on-a-chip devices. Additionally, a-\ch{SiO2} is well suited for chemical surface functionalisation through a variety of techniques, such as physical vapour deposition,\cite{Oldfield2017} electroless plating,\cite{Chen2020} and atomic layer deposition.\cite{Williams2012} Another well-studied technique is the silanisation process: due to the oxide/hydroxyl groups on the surface, alkylsilane or organosilane monolayers can be self-assembled.\cite{Perez-Mitta2019} Those monolayers can then be further modified if necessary. While silane-based surface modifications are possible in other materials like polymers, they require additional steps due to the inability of the functional groups to bind to the surface. These surface modification techniques can potentially enable applications with significantly enhanced performance to currently established techniques. This nanopore system thus provides a new versatile platform for the development of advanced integratable sensor and separation systems.
\begin{acknowledgement}
P.K. acknowledges financial support from the Australian Government through the Australian Research Council under the ARC Discovery Project Scheme (DP180100068). This research was supported by the ANU Grand Challenge Scheme, Our Health in Our Hands (OHIOH). Part of this research was undertaken at the SAXS/WAXS beamline at the Australian Synchrotron, part of ANSTO. The irradiated a-\ch{SiO2} membranes are part of the experiment UMAT, which was performed at the beamline X0 at the GSI Helmhotzzentrum für Schwerionenforschung, Darmstadt (Germany) in the frame of FAIR Phase-0. This work used the ACT node of the NCRIS-enabled Australian National Fabrication Facility (ANFF-ACT).
\end{acknowledgement}

\begin{suppinfo}
The SI is available free of charge: hydrofluoric acid etching setup; scanning electron miscroscopy analysis; atomic force microscopy analysis; small-angle X-ray scattering analysis; pore tip radius calculation; ion transport measurements; $I-V$ data around \SI{0}{\V}.
\end{suppinfo}

\section*{Conflict of Interest}
The authors declare no competing financial interest.

\bibliography{main}

\end{document}